\documentclass[12pt]{article}
\usepackage{times}
\usepackage{cite}
\usepackage{setspace}
\begin{document}
\def\half{\mbox{$\frac{1}{2}$}}
\def\bgk#1{\mbox{\boldmath $#1$}}
\def\bfss#1{{\textsf{\textbf #1}}}
\noindent To be published in {\it Physical Review E} (submitted 6 
October 2015, accepted 6 November 2015)
Abstract:  
https://journals.aps.org/pre/accepted/8607dRbcX4e1be1f42c7857870d7455424a0de39b
\vspace*{1.0in}

\begin{center}
GENERAL FORMALISM FOR SINGLY-THERMOSTATED HAMILTONIAN DYNAMICS\\
~\\~\\
John D. Ramshaw\\~\\Department of Physics\\Portland State University\\
Portland, OR 97207\\

\vspace{0.6in}
ABSTRACT\\

\end{center}

A general formalism is developed for constructing modified Hamiltonian 
dynamical systems which preserve a canonical equilibrium distribution 
by adding a time evolution equation for a single additional thermostat 
variable.  When such systems are ergodic, canonical ensemble averages 
can be computed as dynamical time averages over a single trajectory.  
Systems of this type were unknown until their recent discovery by 
Hoover and colleagues.  The present formalism should facilitate the 
discovery, construction, and classification of other such systems by 
encompassing a wide class of them within a single unified framework.  
This formalism includes both canonical and generalized Hamiltonian 
systems in a state space of arbitrary dimensionality (either even or 
odd), and therefore encompasses both few- and many-particle systems.  
Particular attention is devoted to the physical motivation and 
interpretation of the formalism, which largely determine its structure.  
An analogy to stochastic thermostats and fluctuation-dissipation 
theorems is briefly discussed.\\

\begin{spacing}{2.0}

\newpage
\begin{center}
{\bf I.  INTRODUCTION AND SUMMARY}
\end{center}

In the classical Hamiltonian dynamics of an isolated conservative 
system, the energy is a constant of the motion whose numerical value 
is simply the value of the Hamiltonian at the initial phase point.  
If such a system is ergodic, microcanonical ensemble averages at a 
given fixed energy can be computed as dynamical time averages over 
any single trajectory whose initial phase point lies on that energy 
surface.  More often, however, one would rather compute canonical 
ensemble averages at a given fixed temperature, which implies a 
distribution of energies and a corresponding distribution of initial 
phase points.  Canonical averages therefore cannot be computed as 
dynamical time averages over a single trajectory of the original 
Hamiltonian system.  A great deal of research has been devoted 
to methods for modifying the Hamiltonian dynamics so that an 
entire canonical distribution can be sampled by following a single 
trajectory.  Such modifications or models are commonly referred 
to as {\em thermostats,} which may be broadly characterized as 
stochastic or deterministic.  The former have the advantage that 
they can draw upon the extensive body of methods and results that 
have evolved from the classical Langevin theory of Brownian motion 
\cite{VanK,Mazo,Coffey}.  Deterministic models are simpler 
and more reproducible, but they lack a well-established historical 
foundation analogous to the Langevin theory, so they have had to 
be developed {\em ab initio}.  There was little or no motivation 
to pursue such a development prior to the advent of fast digital 
computers, so most of it has occurred during the past thirty years, 
beginning with the pioneering work of Nos\'{e} \cite{Nose1,Nose2}.

In spite of its relatively brief history, the literature in this 
field has become voluminous.  A bewildering variety of different 
deterministic models has by now been explored with mixed results.  
Fortunately, the most significant subset of this literature has 
now been reviewed in monographs \mbox{\cite{HooverMD,AllenTildesley,
EvansMorriss,HooverCSM,HooverTRCSC,Rapaport,HooversTRCSAC,
HooversSCCNS}}, except of course for very recent developments.  
(The older books cited are now out of date as regards the state 
of the art, but they still contain very useful discussions of 
the fundamentals.)  Our discussion here is therefore restricted 
to papers having direct relevance to the present development.
  
Hoover recognized that Nos\'{e}'s original model was needlessly 
complex, and transformed it into a greatly simplified form now 
referred to as Nos\'{e}-Hoover dynamics \cite{NH}.  Unfortunately, 
it is insufficient for the dynamics to merely preserve a canonical 
equilibrium distribution, because this does not imply that the 
trajectory generated by an arbitrary initial phase point will 
sample the entire distribution.  If it does not, the system is 
not ergodic, time and ensemble averages are not equivalent, and 
the system is not useful.  Alas, ergodicity proofs for systems 
of practical interest are practically nonexistent, so in practice 
ergodicity or its absence must be determined beyond a reasonable 
doubt by numerical experiments.  Such experiments revealed that 
the Nos\'{e} and Nos\'{e}-Hoover models are far from ergodic, and 
therefore cannot be used to compute canonical ensemble averages.

The Nos\'{e}-Hoover model is a singly-thermostated model; i.e., it 
is based on the time evolution of a single additional thermostat 
variable, which plays the role of a linear friction coefficient 
of indefinite sign that can be interpreted as controlling the 
mean kinetic energy of the system.  It was soon discovered that 
ergodicity could be achieved by introducing various combinations 
of {\em nonlinear} frictional and/or kinematic terms involving 
{\em two} thermostat variables \cite{BK,KBB,KB,HH}.  Such models 
then became the state of the art, and remained so for the next 
quarter century or so.  During this period numerous attempts 
were made to construct singly-thermostated ergodic models with 
canonical equilibrium distributions, but a persistent lack of 
success led to a growing suspicion that such models may not exist 
\cite{Japan}.  That suspicion was laid to rest by the surprising 
recent discovery of several such models \cite{HSH,HHS,HSP}, which 
demonstrate by construction that a second thermostat variable is 
not necessary for ergodicity after all.  What does appear to be 
essential is that the friction terms are nonlinear and not too 
simple in form, although they need not be any more complicated 
than those in the doubly-thermostated models \cite{BK,KBB,KB,HH}.  
Our purpose here is to present a compact general formalism which 
encompasses a wide variety of such models by incorporating their 
essential shared features into a unified framework.  A primary 
emphasis is placed on the physical motivation and interpretation 
of the formalism, which largely determine its structure.  The 
resulting general formulation can then readily be specialized 
to generate a wide variety of other similar models, thereby 
facilitating their discovery, development, and classification.  
The formalism includes both canonical and generalized Hamiltonian 
systems in a state space of arbitrary dimensionality $n,$ so that 
it applies to both few- and many-particle systems with either 
integral or half-integral degrees of freedom (i.e., even or odd 
$n$).

The generality of the present treatment should not be allowed 
to obscure the fact that it is based on the same basic physical 
ingredients that underlie most of the previous thermostated 
dynamical models cited above, namely
\vspace{-0.25in}
\end{spacing}

\newcounter{nn}
\begin{list}
{(\alph{nn})}{\usecounter{nn}}
\item The use of generalized frictional terms to produce variations 
in the energy, which would otherwise remain constant.

\item Time-dependent friction coefficients (or ``thermostat 
variables") which alternate between positive and negative values, 
thereby producing alternating periods of decreasing and increasing 
energy, and deterministic energy fluctuations about a nonzero mean 
value.
 
\item Time evolution equations for the friction coefficients, the 
form of which must ensure that the resulting energy distribution 
is canonical with the desired specified temperature.
\end{list}

\begin{spacing}{2.0}
\noindent As will be seen, the structure of the present formalism 
is largely determined by these three essential ingredients.

Our starting point is a generalized formulation of Hamiltonian 
dynamics \cite{Pars,Zwanzig,JDR-KL} which is more compact and 
easier to work with than the canonical form, 
to which it reduces as a special case.  This dynamics conserves 
the energy of the system, so it must be modified to allow the 
energy to vary as required to sample the canonical distribution.  
The natural and obvious physical mechanism by which the energy 
can be varied is friction, which of course is the rationale for 
ingredient (a) above.  It is convenient and natural to introduce 
friction into the generalized Hamiltonian dynamics by means of 
a generalized dissipative term of the form that appears in the 
``mixed canonical and dissipative dynamics" of Enz \cite{Enz}.  
Ingredient (b) is then introduced by affixing a single 
time-dependent friction coefficient to the dissipative term.  
The simplest and most straightforward route to ingredient (c) is 
via the generalized Liouville equation \cite{Liouville,JDRGLE1,
HooverGLE,JDRGLE2}, 
of which the canonical equilibrium distribution is required to 
be a steady-state solution \cite{HooverMD,NH,BK,KBB,KB,SHH}.  
Imposing this requirement leads directly to the required 
time evolution equation for the friction coefficient.  This 
approach was systematically exploited by Kusnezov, Bulgac, 
and Bauer (KBB) \cite{BK,KBB,KB}, who presented a general 
framework for introducing generalized friction terms involving 
two independent thermostat variables into canonical Hamiltonian 
dynamics, and for inferring the time evolution equations those 
variables must satisfy to obtain consistency with a canonical 
equilibrium distribution.  The present development proceeds in 
much the same spirit, but is considerably simpler due to our 
use of generalized Hamiltonian dynamics and a single thermostat 
variable.  Of course, the KBB formulation was developed at a 
time when it was widely suspected that two or more thermostat 
variables are necessary to obtain ergodicity.  In 
light of the recent evidence that a single such variable is 
sufficient \cite{HSH,HHS,HSP}, it seems worthwhile to focus 
attention on singly-thermostated models in greater detail and 
generality.
\newpage
The remainder of the paper is organized as follows.  The equations 
of motion for a generalized Hamiltonian system containing nonlinear 
frictional terms are developed in Sect.~II.  The friction terms 
are taken to be proportional to a single time-dependent scalar 
friction coefficient $z(t).$  A time evolution equation for $z(t)$ 
is derived in Sect.~III by requiring the canonical equilibrium 
distribution to be a stationary solution of the corresponding 
generalized Liouville equation.  In Sect.~IV the general formulation 
is specialized to canonical Hamiltonian dynamics in rectangular 
Cartesian coordinates, which is the most common case of interest.  
In Sect.~V we specialize the canonical formulation to systems with 
a single degree of freedom, which provides a simple direct route 
to the recent models of Hoover et al. \cite{HHS}.  Section VI 
contains a few concluding remarks.\\

\begin{center}
{\bf II.  GENERALIZED HAMILTONIAN DYNAMICS WITH FRICTION}
\end{center}

Our starting point is an arbitrary unmodified Hamiltonian system of 
the general form \cite{Pars,Zwanzig,JDR-KL}
\begin{equation}
\dot{{\bf x}} = {\bfss A}({\bf x}) \cdot {\bgk \nabla} H
\end{equation}
where ${\bf x} = (x_1,x_2,...,x_n)$ is the phase point, $H({\bf x})$ 
is the Hamiltonian function, ${\bgk \nabla} \equiv \partial/\partial 
{\bf x},$ and ${\bfss A}({\bf x})$ is an antisymmetric matrix 
satisfying the condition
\begin{equation}
{\bgk \nabla} \cdot {\bfss A} = {\bf 0}.
\end{equation}
The antisymmetry of ${\bfss A}$ implies at once that $H$ is a constant 
of the motion; i.e., $\dot{H} = \dot{{\bf x}} \cdot {\bgk \nabla} H = 
0.$  Equation (2) combines with the antisymmetry of ${\bfss A}$ to 
imply that ${\bgk \nabla} \cdot ({\bfss A} \cdot {\bgk \nabla} H) = 0,$ 
so that Eq.~(1) generates an incompressible or volume-preserving flow 
in the phase space.  Equation (1) is simpler, more compact, and also 
more general than canonical Hamiltonian dynamics, to which it reduces 
as a special case as discussed in Sect.~IV.

Next we add friction by introducing a dissipative term of the form 
used by Enz \cite{Enz}, so that Eq.~(1) is replaced by
\begin{equation}
\dot{{\bf x}} = {\bfss A}({\bf x}) \cdot {\bgk \nabla} H 
              - \, {\bfss D}({\bf x}) \cdot {\bgk \nabla} H
\end{equation}
where the friction matrix ${\bfss D}({\bf x})$ is symmetric and 
positive semidefinite.  It then follows that 
\begin{equation}
          \dot{H} = - \; {\bgk \nabla} H \cdot {\bfss D} 
                                  \cdot {\bgk \nabla} H \leq 0
\end{equation}
so that the dissipative term produces a monotonic decay in the 
energy, as it was designed to do.  The friction matrix ${\bfss 
D}$ therefore provides provides a convenient general framework 
for introducing artificial terms into the dynamics that possess 
a clear physical interpretation as analogs of real physical 
dissipative effects such as viscosity, friction, and drag.  As 
discussed in the Introduction, however, in the present context 
it is necessary for the energy to alternate between periods of 
growth and decay so that it can generate a canonical energy 
distribution with a nonzero mean value.  The obvious way to 
accomplish this is to simply multiply the dissipative term 
in Eq.~(3) by a time-dependent dimensionless scalar friction 
coefficient $z(t)$ which takes on both positive and negative 
values, so that the dissipation is reversed when $z$ becomes 
negative.  We thereby obtain
\begin{equation}
\dot{{\bf x}} = {\bfss A}({\bf x}) \cdot {\bgk \nabla} H 
         - z \; {\bfss D}({\bf x}) \cdot {\bgk \nabla} H 
                                      \equiv {\bf U}({\bf x},z)
\end{equation}
from which it follows that $\dot{H} = - \; z(t) {\bgk \nabla} 
H \cdot {\bfss D} \cdot {\bgk \nabla} H,$ so that the energy 
increases when $z(t) < 0$ and decreases when $z(t) > 0.$  Since 
ordinary friction coefficients are positive, we shall refer to 
$z(t)$ as a generalized friction coefficient, which serves as 
a reminder that it will alternate between positive and negative 
values, and that it is affixed to a generalized dissipative 
term of the form ${\bfss D} \cdot {\bgk \nabla} H.$  It will 
also be referred to as a (or in this case the) thermostat 
variable, in accordance with common usage in molecular 
dynamics.
  
Although ${\bfss D}$ is normally positive semidefinite in real 
physical systems, one might inquire why we continue to impose 
that requirement in the present context, since multiplying 
${\bfss D}$ by $z$ produces a matrix of indefinite sign.  The 
rationale for retaining this requirement is to preserve as much 
as possible of our physical intuition about the qualitative 
behavior of frictional terms, and to force the variable $z$ to 
carry the entire burden of reversing the sign of such terms.  
However, this is not an absolute requirement, and it would not 
be inconsistent to consider models in which it is relaxed.  
However, if this were done the sign of $\dot{H}$ could no 
longer be determined from $z$ alone.

Of course, Eq.~(5) does not represent the most general possible 
way of introducing a single thermostat variable into the friction 
term.  A much more general form would be obtained by replacing 
$z {\bfss D}({\bf x})$ by ${\bfss F}(z) \cdot {\bfss D}({\bf x})$ 
in Eq.~(5), where ${\bfss F}(z)$ is a symmetric matrix each 
of whose elements is a function of $z.$  However, that level of 
complexity is unmanageable (except in small simple systems) and 
would make it practically impossible to preserve a clear physical 
interpretation of the formalism, and to anticipate the qualitative 
behavior of the dynamics based on physical intuition and insight.
  
One might wonder if it would at least be useful to replace $z$ 
by a scalar function $F(z)$ in Eq.~(5), with the understanding 
that $F(z)$ would likewise need to assume both positive and 
negative values.  This apparent additional generality would 
be specious and illusory, however, since $\dot{F} = (dF/dz) 
\, \dot{z},$ so the time evolution of $z$ defines \pagebreak 
and determines the time evolution of $F(z),$ and vice versa.  
Specifying the former is therefore equivalent to specifying 
the latter, so replacing $z$ by $F(z)$ is tantamount to simply 
denoting $z$ by the different symbol $F.$\\

\begin{center}
{\bf III.  THE TIME EVOLUTION OF THE FRICTION COEFFICIENT}
\end{center}

The time evolution of $z$ is not known or specified {\it a 
priori,} so it must be determined by considering $z$ as an 
additional dynamical variable.  The enlarged state space of the 
system then becomes $({\bf x},z),$ and in order for the 
system to remain deterministic and autonomous $z$ must obey 
a time evolution equation of the form
\begin{equation}
\dot{z} = W({\bf x},z)
\end{equation}
which together with Eq.~(5) determines the time evolution of the 
system.  It is now necessary to determine what form the function 
$W({\bf x},z)$ must have in order for the dynamics to 
generate a canonical distribution in the energy.  Fortunately, 
this need not be done by trial and error but can be accomplished 
in a systematic way simply by requiring a canonical distribution 
with a specified temperature to be a stationary solution of the 
generalized Liouville equation \cite{Liouville,JDRGLE1,HooverGLE,
JDRGLE2}, which in the present context takes the form
\begin{equation}
\frac{\partial \rho}{\partial t} 
       + {\bgk \nabla} \cdot (\rho {\bf U}) 
       + \frac{\partial(\rho W)}{\partial z} = 0
\end{equation}
where $\rho({\bf x},z,t)$ is the normalized probability 
distribution in the variables $({\bf x},z)$ at time $t.$ 
The form of $W$ is constrained by the requirement that 
Eq.~(7) must possess a steady (time-independent) solution 
$\rho_s({\bf x},z)$ whose reduced distribution in ${\bf x}$ 
alone is canonical; i.e., \pagebreak 
\begin{equation}
\int dz \; \rho_s({\bf x},z) = \rho_c({\bf x}) 
\equiv Q^{-1} \exp\{-\beta H({\bf x})\}
\end{equation}
where $\beta = 1/T,$ $T$ is the specified temperature 
in energy units, and $Q(\beta) \equiv \int d{\bf x} \, 
\exp\{-\beta H({\bf x})\}$ is the canonical partition function.  
The simplest way to satisfy Eq.~(8) is to require that ${\bf x}$ 
and $z$ be statistically independent in steady state, so that
\begin{equation}
\rho_s({\bf x},z) = \rho_c({\bf x}) \; \sigma(z)
\end{equation}
where $\sigma(z)$ is nonnegative and satisfies $\int dz \, 
\sigma(z) = 1,$ but is otherwise as yet undetermined.

Combining Eqs.~(5) and (7)--(9), we obtain, after a little algebra,
\begin{equation}
{\bgk \nabla} \cdot ({\bfss D} \cdot {\bgk \nabla} H) 
    - \beta {\bgk \nabla} H \cdot {\bfss D} \cdot {\bgk \nabla} H 
       = \frac{1}{z \sigma} \frac{\partial (\sigma W)}{\partial z}
\end{equation}
where use has been made of Eq.~(2) and the antisymmetry of ${\bfss 
A}.$  Even if $\sigma(z)$ were given or known, Eq.~(10) obviously 
does not uniquely determine $W,$ because it leaves $\sigma W$ 
undetermined to within an arbitrary function of ${\bf x}.$  This 
nonuniqueness affords us the freedom to restrict attention to 
functions $W({\bf x},z)$ of the separable form
\begin{equation}
W({\bf x},z)  = \varphi(z) \chi({\bf x})
\end{equation}
so that Eq.~(6) becomes
\begin{equation}
\dot{z} = \varphi(z) \chi({\bf x})
\end{equation}
The function $\varphi(z)$ provides a mechanism whereby the rate 
at which the generalized friction coefficient $z(t)$ responds 
to ${\bf x}$ can be either accelerated or retarded based on the 
value of $z$ itself, which is a potentially useful feature.  
It is clear, however, that $\varphi(z)$ should not be allowed 
to change the sign of $\dot{z},$ since whether $z$ needs to 
increase or decrease to maintain a canonical distribution in 
${\bf x}$ is determined entirely by ${\bf x}.$ Thus $\varphi(z)$ 
cannot change sign, and we can require $\varphi(z) > 0$ with no 
loss in generality.  Equation (10) now becomes  
\begin{equation}
\frac{1}{\chi({\bf x})} [{\bgk \nabla} \cdot ({\bfss D} \cdot 
{\bgk \nabla} H) - \beta {\bgk \nabla} H \cdot {\bfss D} \cdot 
{\bgk \nabla} H] = \frac{1}{z \sigma} \frac{d(\sigma \varphi)}{dz}
\end{equation}

The left member of Eq.~(13) depends only on ${\bf x},$ while the 
right member depends only on $z,$ so they must both have the same 
constant value, which we denote by $- 1/\alpha.$  It follows that
\begin{equation}
\chi({\bf x}) = \alpha[\beta {\bgk \nabla} H \cdot {\bfss D} 
\cdot {\bgk \nabla} H - {\bgk \nabla} \cdot ({\bfss D} \cdot 
{\bgk \nabla} H)]
\end{equation}
and
\begin{equation}
\alpha \frac{d(\sigma \varphi)}{dz} = -  \; z \sigma
\end{equation}
Equation (15) implies that $\alpha$ must be positive, for if it 
were negative $\sigma$ would not in general be normalizable, as 
shown by the special case $\varphi = 1,$ in which the solution of 
Eq.~(15) is $\sigma(z) = \sigma(0) \exp\{- z^2/(2\alpha)\}.$  It 
follows that the factors of $\alpha$ in Eqs.~(14) and (15) can 
simply be absorbed into the positive quantities ${\bfss D}$ and 
$\varphi,$ which is formally tantamount to simply setting $\alpha 
= 1.$  Equations (14) and (15) then simplify to
\begin{equation}
\chi({\bf x}) = \beta {\bgk \nabla} H \cdot {\bfss D} 
\cdot {\bgk \nabla} H - {\bgk \nabla} \cdot ({\bfss D} \cdot 
{\bgk \nabla} H)
\end{equation}
\begin{equation}
\frac{d(\sigma \varphi)}{dz} = -  \; z \sigma 
\end{equation}

Equation (16) explicitly determines $\chi({\bf x}),$ while Eq.~(17) 
is an ordinary differential equation which relates $\sigma(z)$ and 
$\varphi(z).$  One of those functions can be chosen at will (subject 
to the aforementioned constraints) and Eq.~(17) then determines the 
other.  The functional form of $\sigma(z)$ is essentially immaterial, 
since it has no effect on the canonical equilibrium distribution 
$\rho_c({\bf x}),$ whereas $\varphi(z)$ directly affects the dynamics 
via Eq.~(12).  It is therefore natural to regard $\varphi(z)$ as a 
specified function, which then determines $\sigma(z)$ via Eq.~(17).  
Equation (17) can be rewritten as
\begin{equation}
\frac{d}{dz}\log(\sigma \varphi) = -  \; \frac{z}{\varphi}
\end{equation}
which can immediately be integrated to obtain
\begin{equation}
\sigma(z) = \frac{C}{\varphi(z)} 
           \exp\left\{-\int_0^z \frac{s \, ds}{\varphi(s)} \right\}
\end{equation}
where $C$ is a constant of integration.  Equation (19) provides an 
explicit expression for $\sigma(z)$ for a given $\varphi(z) > 0.$  
The constant $C$ is not arbitrary or unknown, because it is simply 
determined by the normalization condition $\int dz \, \sigma(z) = 
1.$  In the special case $\varphi(z) = 1,$ Eq.~(19) reduces to 
\begin{equation}
\sigma(z) = C \exp \{- \half z^2\}
\end{equation}
where $C = 1/\sqrt{2\pi},$ so the steady-state distribution in $z$ 
is Gaussian in this case, with a variance of $\langle z^2 \rangle 
= \int dz \, z^2 \sigma(z) = 1.$

\newpage
According to Eqs.~(12) and (16), $\dot{z}$ is proportional to 
${\bfss D},$ and as previously noted $\dot{H} = - \; z(t) {\bgk 
\nabla} H \cdot {\bfss D} \cdot {\bgk \nabla} H$ is likewise 
proportional to ${\bfss D}.$  It follows that the time scales 
for the dynamical evolution of both $z$ and $H$ can be made 
shorter simply by making ${\bfss D}$ larger (e.g., multiplying 
it by a dimensionless scalar coefficient $> 1$), which would 
therefore be expected to produce a faster approach to the 
steady-state distribution and faster asymptotic convergence of 
time averages.  Beyond a certain point, however, shorter time 
scales result in stiffness and numerical inefficiency, and in 
any case it seems pointless to make the time scales associated 
with ${\bfss D}$ any shorter than those associated with ${\bfss 
A}$ in the original Hamiltonian system of Eq.~(1), which also 
limit the rate of approach to a steady-state distribution.  
Ideally it would be preferable for all the various time scales 
to be of the same order of magnitude, so in practice it may be 
advantageous to rescale or renormalize ${\bfss D}$ to that end, 
which will require numerical experimentation.

The general formalism is now complete.  The dynamical evolution 
of the phase point $({\bf x},z)$ is determined by Eqs.~(5) and 
(12), in which $\chi({\bf x})$ is given by Eq.~(16).  The 
function $\varphi(z) > 0$ and the positive semidefinite matrix 
${\bfss D}$ may be chosen at will.  By construction, any dynamical 
system of this form will preserve a steady-state probability 
distribution of the canonical form (9), where $\rho_c({\bf x})$ 
and $\sigma(z)$ are given by Eqs.~(8) and (19), and if the 
system is ergodic it will dynamically generate that distribution 
as $t \rightarrow \infty$ starting from almost all initial phase 
points $({\bf x}_0,z_0).$\\

\begin{center}
{\bf IV.  CANONICAL HAMILTONIAN SYSTEMS}
\end{center}

Here we specialize the general formalism to canonical Hamiltonian 
systems with $f$ degrees of freedom, for which $n = 2f$ is even.  
For a system of $N$ point particles in $d$-dimensional space, $f 
= N d.$  We now have ${\bf x} = ({\bf q},{\bf p})$ and ${\bgk 
\nabla} H = (\partial H/\partial{\bf q},\partial H/\partial 
{\bf p}),$ where ${\bf q} = (q_1,,...,q_f)$ and ${\bf p} = 
(p_1,,...,p_f)$ are the canonical coordinates and momenta.  The 
canonical form of the matrix ${\bfss A}$ is 
\begin{equation}
{\bfss A} = 
\left[\begin{array}{cc}
  {\bfss 0} & {\bfss 1} \\
- {\bfss 1} & {\bfss 0}
\end{array}\right]
\end{equation}
where ${\bfss 0}$ and ${\bfss 1}$ are the zero and unit $f \times 
f$ matrices, respectively.  It then follows that ${\bfss A} \cdot 
{\bgk \nabla} H = (\partial H/\partial{\bf p},- \partial H/\partial 
{\bf q}),$ whereupon Eq.~(1) immediately yields the canonical 
Hamiltonian equations of motion $\dot{\bf q} = \partial H/\partial 
{\bf p}$ and $\dot{\bf p} = - \, \partial H/\partial{\bf q}.$

We shall restrict attention to conservative systems in rectangular 
Cartesian coordinates, in which $H({\bf q},{\bf p})$ assumes the 
familiar form
\begin{equation}
H({\bf q},{\bf p}) = \frac{1}{2}|{\bf p}|^2 + V({\bf q})
\end{equation}
where $V({\bf q})$ is the potential energy, and all partlcle 
masses have been set equal to unity for simplicity.  It follows 
that $\partial H/\partial{\bf q} = \partial V/\partial{\bf q}$ 
and $\partial H/\partial{\bf p} = {\bf p},$ so that ${\bgk 
\nabla} H = (\partial V/\partial{\bf q},{\bf p})$ and Eq.~(1) 
immediately reduces to the Newtonian equations of motion
\begin{eqnarray}
\dot{{\bf q}} &=& {\bf p}\\
\dot{{\bf p}} &=& - \, \frac{\partial V}{\partial {\bf q}}
\end{eqnarray}

In the present context, the friction matrix ${\bfss D}$ is of 
the general form
\begin{equation}
{\bfss D}({\bf q},{\bf p}) = \left[
\begin{array}{cc}
{\bfss D}_q({\bf q},{\bf p}) & 
                       {\bfss D}_\triangle({\bf q},{\bf p})\\
{\bfss D}_\triangle^{{\sf T}}({\bf q},{\bf p}) & 
                       {\bfss D}_p({\bf q},{\bf p})
\end{array} \right]
\end{equation}
where ${\bfss D}_q$ and ${\bfss D}_p$ are symmetric, whereas 
${\bfss D}_\triangle$ need not be, and superscript ${\sf T}$ 
denotes the transpose.  According to Eq.~(5), the matrices 
${\bfss D}_q$ and ${\bfss D}_\triangle$ will have the effect 
of introducing generalized frictional terms into Eq.~(23).  
Such terms have indeed frequently been employed in previous 
thermostat models (e.g., \cite{KBB}), but they seem undesirable 
if not hazardous on the grounds that Eq.~(23) is essentially 
kinematical in nature, so tampering with it is conceptually 
incongruous and may tend to confound our physical intuition.  
To forestall this danger, we shall simply set ${\bfss D}_q = 
{\bfss D}_\triangle = {\bfss 0},$ so that Eq.~(25) reduces to
\begin{equation}
{\bfss D}({\bf q},{\bf p}) = \left[
\begin{array}{cc}
{\bfss 0} & {\bfss 0}\\
{\bfss 0} & {\bfss D}_p({\bf q},{\bf p})
\end{array} \right]
\end{equation}
It then follows that ${\bfss D} \cdot {\bgk \nabla} H = 
({\bfss 0},{\bfss D}_p \cdot {\bf p}),$ so that Eq.~(5) 
reduces to
\begin{eqnarray}
\dot{\bf q} &=& {\bf p}\\
\dot{\bf p} &=& - \, \frac{\partial V}{\partial {\bf q}} 
      - \, z \, {\bfss D}_p \cdot {\bf p}
\end{eqnarray}
in which it is obvious that the term involving ${\bfss D}_p$ 
is a friction term, which is generally nonlinear due to the 
dependence of ${\bfss D}_p$ on $({\bf q},{\bf p}).$  This 
term still preserves a great deal of generality, since it 
allows for frictional couplings between different degrees 
of freedom.  This is probably much more generality than is 
necessary for practical purposes, so we shall adopt the 
further simplification that ${\bfss D}_p$ is diagonal with the 
form ${\bfss D}_p = {\textbf [} \delta_{ij} \gamma_i(q_i,p_i) 
{\textbf ]},$ where ${\textbf [} a_{ij} {\textbf ]}$ denotes 
the matrix whose $(i,j)$ element is $a_{ij},$ $\delta_{ij}$ 
is the Kroencker delta, and $\gamma_i \geq 0$ so that 
${\bfss D}_p$ is positive semidefinite.  Equation~(28) then 
becomes \pagebreak
\begin{equation}
\dot{p}_i = - \, \frac{\partial V}{\partial q_i} 
                  - z \, \gamma_i(q_i,p_i) \, p_i
\end{equation}
Notice that this form still preserves the dependence of the 
friction term on $q_i,$ and thereby provides the option to 
localize the thermostating in coordinate space.  Thus, for 
example, it could be restricted to regions at or near the 
boundaries of a finite region.

All that remains is to specialize the equation of motion for 
$z(t)$ to the present situation, which merely requires us to 
evaluate the quantities in Eq.~(16) and combine the result 
with Eq.~(12).  We readily find that 
\begin{equation}
{\bgk \nabla} H \cdot {\bfss D} \cdot {\bgk \nabla} H = 
    {\bfss D}_p : {\bf p}{\bf p} = \sum_i \gamma_i \, p_i^2
\end{equation} 
and
\begin{equation}
{\bgk \nabla} \cdot ({\bfss D} \cdot {\bgk \nabla} H) = 
\frac{\partial}{\partial {\bf p}} \cdot ({\bfss D}_p \cdot 
{\bf p}) = \sum_i \frac{\partial(\gamma_i p_i)}{\partial p_i}
\end{equation}
Equations (12), (16), (30), and (31) then combine to yield
\begin{equation}
\dot{z} = \varphi(z) \sum_i \left(\beta \gamma_i \, p_i^2 
      - \frac{\partial(\gamma_i p_i)}{\partial p_i}\right)
\end{equation}
in which $\varphi(z) > 0$ can be chosen at will.  We emphasize 
that the number of degrees of freedom $f$ remains arbitrary, 
so these equations apply to both small and many-particle 
Hamiltonian systems.  It is noteworthy that for this class 
of models, the time evolution equation (32) for $z(t)$ is 
entirely independent of the form of the potential $V({\bf 
q}).$\\

\newpage
\begin{center}
{\bf V.  SYSTEMS WITH A SINGLE DEGREE OF FREEDOM}
\end{center}

A considerable portion of the previous work on thermostated 
dynamics has restricted attention to systems with a single 
degree of freedom, both because of their intrinsic theoretical 
interest and in order to gain insight into various alternative 
types of thermostated dynamics in situations which are more 
easily comprehensible and less computationally demanding.  It 
is therefore of interest to specialize the above development 
to systems with only a single degree of freedom, for which 
${\bf x} = (q,p)$ and $H = \half p^2 + V(q).$  Equations (27), 
(29), and (32) then immediately reduce to
\begin{eqnarray}
\dot{q} &=& p\\
\dot{p} &=& - \, \frac{dV}{dq} - z \gamma p\\
\dot{z} &=& \varphi(z) \left(\beta \gamma p^2 
           - \frac{\partial(\gamma p)}{\partial p}\right)
\end{eqnarray}
As previously noted, Eq.~(35) is is independent of the form of 
$V(q).$  For simplicity we further restrict attention to models 
in which $\gamma = \gamma(p)$ is independent of $q,$ so that 
Eq.~(35) reduces to
\begin{equation}
\dot{z} = \varphi(z) \left(\beta \gamma p^2 
                              - \frac{d(\gamma p)}{dp}\right)
\end{equation}
The simplest choices for $\varphi(z)$ and $\gamma(p)$ are 
$\varphi = 1$ and a constant $\gamma$ independent of $p.$  
With those choices, Eq.~(36) becomes
\begin{equation}
\dot{z} = \gamma (\beta p^2 - 1)
\end{equation}
For a one-dimensional simple harmonic oscillator with \pagebreak 
$V = \half q^2,$ Eqs.~(33), (34), and (37) reduce to the 
Nos\'{e}-Hoover model \cite{NH} (with $\zeta = \gamma z$).  It 
is reassuring to see this model emerge so easily as a simple 
special case of the general formalism.

Numerous detailed studies have confirmed that the Nos\'{e}-Hoover 
model is not ergodic; it is evidently insufficiently nonlinear.  
Subsequent studies showed that friction terms nonlinear in $p$ are 
conducive to ergodicity \cite{KBB,HSH,HHS,SHH}.  Since $\gamma(p) 
> 0,$ the simplest such terms are obtained by letting $\gamma(p)$ 
be a low-order polynomial in the dimensionless variable $\beta 
p^2$ with positive coefficients.  Thus we are led to consider
\begin{equation}
\gamma(p) = a + b \beta \, p^2 + c \beta^2 \, p^4
\end{equation}
which combines with Eqs.~(34) and (36) to yield
\begin{eqnarray}
\dot{p} &=& - \, \frac{dV}{dq} 
                  - z(a p + b \beta \, p^3 + c \beta^2 \, p^5)\\
\dot{z} &=& \varphi(z) [a(\beta p^2 - 1) + b(\beta^2 p^4 - 3 
                   \beta p^2) + c (\beta^3 p^6 - 5 \beta^2 p^4)]
\end{eqnarray}
If we transform variables by letting $z = \zeta^\nu$ (where $\nu$ 
is a positive integer which must be odd so that $z$ can change 
sign) and set $\varphi(z) = dz/d\zeta = \nu \zeta^{\nu - 1} = 
\nu z^{1-1/\nu},$ Eqs.~(39) and (40) become
\begin{eqnarray}
\dot{p} &=& - \, \frac{dV}{dq} 
         - \zeta^\nu(a p + b \beta \, p^3 + c \beta^2 \, p^5)\\
\dot{\zeta} &=& a(\beta p^2 - 1) + b(\beta^2 p^4 - 3 
                   \beta p^2) + c (\beta^3 p^6 - 5 \beta^2 p^4)
\end{eqnarray}
which is precisely a model recently discovered and analyzed by 
Hoover, Hoover, \& Sprott (HHS) \cite{HHS} for the cases of a 
harmonic oscillator ($V = \half q^2$) and pendulum ($V = - \cos 
q$), in both of which the model appears to be ergodic for certain 
values of $(a,b,c).$  The steady-state probability distribution 
in $\zeta$ is easily obtained by using Eq.~(19) to evaluate 
$\sigma(z)$ and transforming it to the variable $\zeta$ (which 
of course requires multiplication by the Jacobian $dz/d\zeta$).  
When this is done the result is found to be 
$C \exp\{- \zeta^{\nu+1}/(\nu+1)\},$ in agreement with HHS.  It 
is satisfying to see how the present formalism leads naturally 
to the HHS model in a very straightforward way.

During the evolution of thermostated dynamical models over 
the past thirty years, it has been found useful and insightful 
to interpret thermostat variables like $z$ as integral control 
variables, and to interpret terms of the form appearing 
in Eq.~(40) as controling the moments of the probability 
distribution \cite{HooverMD,HooversSCCNS,HH,HSH,HHS}.  In the 
present development, however, such interpretations are merely 
ancillary and no longer guide the construction of the models, 
since the dependence of $\dot{z}$ on ${\bf x}$ is entirely 
determined by the assumed form of the friction terms via 
Eqs.~(12) and (16).  There seems no {\it a priori} reason to 
restrict attention to frictional terms of polynomial form, so 
it may be worthwhile to explore the behavior of other nonlinear 
functions; e.g., $\gamma(p) = a \cosh(b \sqrt{\beta} \, p).$\\

\begin{center}
{\bf VI.  CONCLUDING REMARKS}
\end{center}

The generalized Hamiltonian formalism on which the present 
development is based was previously used as a framework for 
introducing multiplicative noise and nonlinear dissipation 
into Hamiltonian dynamics in such a way as to preserve a 
canonical equilibrium distribution [25].  The resulting 
nonlinear Langevin equation can be regarded as a stochastic 
thermostat, of which the present development constitutes 
a close deterministic analog.  In the Langevin theory, 
the noise and dissipation are represented by separate 
additive terms with nonlinear coefficients, which are 
related by a fluctuation-dissipation theorem.  In the 
present development, the deterministic function $z(t)$ 
plays somewhat the same role as the random noise in the 
stochastic theory, but is combined with the friction into 
a single purely multiplicative term $z(t) \; {\bfss D}
\cdot {\bgk \nabla} H$ of indefinite sign.  Thus the 
stochastic and deterministic formulations differ somewhat 
in structure, but they share the common feature that 
their respective time-dependent forcing functions are 
self-consistently determined by requiring the resulting 
steady-state probability distribution in ${\bf x}$ to be 
canonical.  In the stochastic theory that requirement 
leads to the nonlinear fluctuation-dissipation theorem, 
while in the deterministic theory it leads to the relation 
between $z(t)$ and ${\bfss D}$ given by Eqs.~(12) and (16).  
The latter relation therefore constitutes a deterministic 
analog of the stochastic fluctuation-dissipation theorem, 
so it seems natural to regard and refer to it as a {\it 
deterministic fluctuation-dissipation} (DFD) {\it theorem.}

Finally, we remark that the general formalism of Sects.~II 
and III could be extended if desired, at the cost of some 
additional complexity, to accommodate multiply-thermostated 
models with two or more friction coefficients $(z_1(t),z_2(t),
\cdots).$  This could be done simply by inserting factors of 
$z_\mu$ ($\mu = 1, 2, \cdots$) into the individual elements 
of the matrix ${\bfss D}$ in Eq.~(3), but it seems preferable 
to first determine the orthogonal matrix ${\bfss M}({\bf x})$ 
which diagonalizes ${\bfss D},$ so that ${\bfss D} = {\bfss 
M} \cdot {\textbf [} \gamma_i \delta_{ij} {\textbf ]} \cdot 
{\bfss M}^{{\sf T}},$ where the $\gamma_i({\bf x})$ are the 
nonnegative eigenvalues of ${\bfss D}.$  Any or all of the 
nonzero $\gamma_i({\bf x)}$ could then be multiplied by 
factors of $z_\mu$ as desired, and the result combined with 
Eq.~(3) to obtain a multiply-thermostated generalization of 
Eq.~(5).\\

\begin{center}
{\bf ACKNOWLEDGMENTS}
\end{center}

I am indebted to Bill Hoover for rekindling my interest in 
this area, for many stimulating and enlightening discussions, 
and for generously sharing his preprints and new results prior 
to publication.

\vspace{-0.2in}
\parindent 0in

\end{spacing}
\newpage
\begin{center}
{\bf REFERENCES\\~}
\end{center}


\begingroup
\renewcommand{\section}[2]{}

\endgroup
\end{document}